\documentclass{kluwer}    
\usepackage[dvips]{graphicx}

\def\cf{{cf.~}}
\def\eg{{e.g.~}}

\begin{document}                                                                                   
\begin{article}
\begin{opening}         
\title{General Relativistic Magnetoionic Theory}
\author{Avery \surname{Broderick} \& Roger \surname{Blandford}}  
\runningauthor{Avery Broderick \& Roger Blandford}
\runningtitle{General Relativistic Magnetoionic Theory}
\institute{Theoretical Astrophysics, Caltech, Pasadena, CA 91125, USA}
\date{August 24, 2002}

\begin{abstract}
We have developed methods for tracing rays and performing radiative
transfer through a magnetoactive plasma in a general relativistic
environment.  The two electromagnetic plasma modes propagate
differently due to a combination of dispersive and gravitational
effects.  We have found that, when given an appropriate environment
surrounding the central black hole, it is indeed possible to generate a
significant degree of circular polarisation without an appreciable
amount of linear polarisation due to these effects alone.
\end{abstract}

\end{opening}           

\section{Introduction}

Polarisation measurements now exist for many accreting compact
objects (ostensibly black holes) at a number of frequencies.
Typically, emission mechanisms are called upon to explain polarisation
observations (see \eg \opencite{Brom-Meli-Liu:01};~\opencite{West:59}).
However, recent
observations of Sgr A$^*$ and M81 (see \eg \opencite{Brun-Bowe-Falc-Mell:01}
;~\opencite{Bowe-Falc-Back:99};~
\opencite{Saul-Macq:99}), as well
as a number of blazars (see \eg \opencite{Kome-Robe-Miln-Rayn-Cook:84}),
have exhibited a significant amount
of circular polarisation (CP) apparently unassociated with any linear
polarisation (LP).  This has proved difficult to explain with the standard
set of polarised emission mechanisms alone, often requiring
specialised magnetic field or disk structures.  In addition to its
anomalous size, the CP typically does not
change in sign despite having a variability larger in frequency and
magnitude than that of the LP (if present) and total
intensity.  Both of these suggest that the region  responsible for the
polarisation is compact, and perhaps the central  compact
object is playing a significant role, if only in moderating the local
plasma and/or magnetic field structure.  As a result, a substantial effort
has been made to investigate the effects of the accretion environment
upon polarisation.

These efforts have been primarily concentrated in two directions: ({\em i})
propagation effects due to a magnetised plasma (see \eg \opencite{Rusz-Bege:02}
;~\opencite{Macq:02};~\opencite{Jone-ODel:77a} \& \citeyear{Jone-Odel:77b}),
and ({\em ii}) vacuum propagation
effects due to general relativity, in particular near a rotating black
hole (see \eg \opencite{Falc-Meli-Agol:00};~\opencite{Agol:97};~
\opencite{Laor-Netz-Pira:90};~\opencite{Conn-Star-Pira:80}).  Most of these
require an initial source of polarisation, presumably provided by the
emission mechanism.  A notable exception is the scintillation
mechanism proposed by \citeauthor{Macq-Melr:00}.  However, for realistic
conditions this has been unable to produce a polarisation of constant
sign.  The studies regarding ({\em i}) have thus far ignored general
relativistic effects (and hence are inapplicable near the compact object),
focusing upon non-dispersive plasma effects, \eg Faraday rotation and
conversion.
The work considering ({\em ii}) has found general
relativity to have a depolarising influence on LP due to frame dragging
for photons passing near the black hole. 
However, the studies of general relativistic effects
have ignored plasma effects completely, and hence are not always
applicable in the case of a thick disk or when a dense and/or
magnetised corona is present.

In contrast, magnetoionic effects, including dispersion, have been
studied in detail in the context of radio waves in the upper
atmosphere.  This has, of course, been done in the absence of general
relativity, where it has been found that for a specific range in
frequency the dispersive effects can have a significant impact
upon the propagation and polarisation of the radio waves (see
\eg \opencite{Budd:64}).

Here we present a fully general relativistic magnetoionic theory which
takes into account general relativity as well as dispersive and
non-dispersive plasma effects.  This is a natural, albeit
currently less well developed, extension of the previous
investigations into the polarisation effects of accretion flows onto
compact objects.  The development of the theory can be succinctly
separated into the problems of tracing rays and performing the
radiative transfer.  As such, these proceedings will be presented in five
sections with \S2 discussing ray tracing, \S3 explaining the radiative
transfer, \S4 presenting results for Bondi flows, and \S5 containing
conclusions.

\section{Ray Tracing}

\subsection{Formalism}

The appropriate place to begin the study of photon propagation in a plasma
are Maxwell's equations,
\begin{equation}
\nabla_\mu F^{\nu \mu} = 4 \pi J^\nu
\;\; \mbox{and} \;\;
\nabla_\mu \mbox{}^*\!F^{\nu \mu} = 0 \,,
\label{maxwell1}
\end{equation}
here expressed in covariant form in terms of the electromagnetic field tensor,
$F^{\nu \mu} \equiv \nabla^\nu A^\mu - \nabla^\mu A^\nu$,
its dual $^*\!F^{\nu \mu}$, and the current fourvector, $J^\nu$.
A prescription is required to determine the current from the electromagnetic
field.  For small fields, this may be accomplished by a 
covariant extension of Ohm's law,
\begin{equation}
J^\nu = \sigma^\nu_{~\mu} E^\mu
\;\;\mbox{where}\;\;
E^\mu \equiv F^{\mu \nu} u_\nu
\label{ohms_law}
\end{equation}
is the fourvector coincident with the electric field vector in the locally flat
comoving rest (LFCR) frame of the plasma ($u^\nu$ is the average plasma
velocity fourvector).
Inserting Ohm's law into Maxwell's equations and expressing the result in
terms of $E^\mu$ and $B^\mu \equiv~^*\!F^{\mu \nu} u_\nu$, yields
\begin{eqnarray}
\nabla_\mu  \left( u^\nu E^\mu - E^\nu u^\mu + 
\varepsilon^{\nu\mu\alpha\beta}\,u_\alpha\,B_\beta \right) &=& 
4 \pi \sigma^\nu_{~\mu} E^\mu \label{maxwell2a} \\
\nabla_\mu  \left( u^\nu B^\mu - B^\nu u^\mu + 
\varepsilon^{\nu\mu\alpha\beta}\,u_\alpha\,E_\beta \right) &=& 
0 \label{maxwell2b} \,,
\end{eqnarray}
where $\varepsilon^{\nu\mu\alpha\beta}$ is the Levi-Civita pseudo tensor.
These are eight partial differential equations which may be solved 
for $E^\mu$ and $B^\mu$ given a specific form
for the conductivity tensor, $\sigma^\nu_{~\mu}$.

Solving these equations can be greatly simplified by making use of a two
length scale expansion (the so-called WKB or Eikonal approximations).  This
is permitted because the photon wavelengths of interest are much smaller than
both, the typical general relativistic length scale (the size of the black
hole), and the typical plasma scale length.  In covariant form
this expansion takes the form of assuming that $E^\mu$ and $B^\mu$ are 
proportional to a phase factor $\exp{\left(iS\right)}$ where the action,
$S$, is related to the wave fourvector by $k_\mu = \nabla_\mu S$.  Keeping
only the lowest order terms and combining equations (\ref{maxwell2a}) and 
(\ref{maxwell2b}) gives
\begin{equation}
\left( k^\delta k_\delta g_{\mu \nu} + k_\mu k_\nu 
+ 4 \pi i \omega \sigma_{\mu \nu} \right) E^\nu = 0 \,,
\label{wave_eq}
\end{equation}
where $\omega \equiv k_\mu u^\mu$ is the photon frequency in the LFCR frame.
From this equation it is possible to
determine the polarisation (for
conductivities with non-degenerate polarisation eigenmodes) and a dispersion
relation, $D(x^\mu, k_\mu)$, a scalar function of the position and wave 
fourvectors that vanishes along a ray.  From the latter it is possible to
construct the rays directly using a covariant extension of the
Hamilton-Weinberg equations (\cf \opencite{Wein:62}),
\begin{equation}
\frac{d x^\mu}{d \lambda} = \frac{\partial D}{\partial k_\mu}
\;\;\mbox{and}\;\;
\frac{d k_\mu}{d \lambda} = - \frac{\partial D}{\partial x^\mu} \;,
\label{hamiltons_eqs}
\end{equation}
where $\lambda$ is an affine parameter, the details of which depend upon the
precise form of $D$ chosen.

\subsection{Dispersion Relations}

In order to investigate the implications of the ray equations
(equations (\ref{hamiltons_eqs})) it is instructive to consider a number
of particular dispersion relations.  First, consider that corresponding to
de Broglie waves, or particles,
\begin{equation}
D = k^\mu k_\mu + m^2 \,.
\label{deBroglie}
\end{equation}
When inserted into the ray equations this produces the
geodesic equations, corresponding to test particles in general relativity.

Second, consider the dispersion relation associated with an isotropic plasma
(derived from equation (\ref{wave_eq}) with the appropriate conductivity,
\cf \opencite{Kuls-Loeb:92}),
\begin{equation}
D = k^\mu k_\mu + \omega_P^2 \,,
\label{isoplasma}
\end{equation}
where $\omega_P^{} \equiv \sqrt{ 4 \pi e^2 n_e / m_e}$ is the plasma frequency
and $n_e$ is the proper electron density.  This bears a striking resemblance
to equation (\ref{deBroglie}), with the plasma frequency taking the place
of a mass.  Hence photons in a plasma act as if they have mass,
with one significant difference: now this ``mass'' depends upon position
through the plasma density.  Therefore, in general photons in a plasma will
not follow geodesics, and in particular will not follow the null geodesics
that photons follow in vacuum.

Third, consider the dispersion relation associated with an magnetoactive
plasma in the quasi-longitudinal approximation (again this is derived
from equation (\ref{wave_eq})),
\begin{equation}
D = k^\mu k_\mu + \frac{\omega \omega_P^2}{\omega \pm \omega_B^{}} \,,
\end{equation}
where $\omega_B^{} \equiv e \sqrt{{\cal B}^\mu {\cal B}_\mu } / m_e$
is the cyclotron frequency associated with the externally imposed magnetic
field, ${\cal B}^\mu$, and the $\pm$ runs over the two different polarisation
eigenmodes. Again, this is similar to equation (\ref{deBroglie}) with the
exception that now the ``mass'' depends upon polarisation as well as
position.  As a result, the different polarisation modes will propagate along
different paths. This is simply an expression of the dispersive nature of
magnetoactive plasmas.

Finally, for completeness the general dispersion relation for the
magnetoactive, cold electron plasma is given by
\begin{eqnarray}
D &=& k^\mu k_\mu - \delta \omega^2 
- \frac{\delta}{2\left(1+\delta\right)} 
\Bigg\{ \left[ \left( \frac{e {\cal B}^\mu k_\mu}{m_e \omega} \right)^2 
- \left(1+2\delta \right) \omega_B^2 \right] \nonumber \\
&\pm& \left. \sqrt{\left( \frac{e {\cal B}^\mu k_\mu}{m_e \omega} \right)^4 +
2 \left(2\omega^2-\omega_B^2-\omega_P^2\right) 
\left( \frac{e {\cal B}^\mu k_\mu}{m_e \omega} \right)^2 
+ \omega_B^4} \right\} \\
&&\mbox{where} \;\; \delta \equiv \frac{\omega_P^2}{\omega_B^2-\omega^2} \,.
\nonumber
\end{eqnarray}
This is the covariant extension of the Appleton-Hartree dispersion relation
(see \eg \opencite{Budd:64} or \opencite{Boyd-Sand:69}).  The extension
to a pair plasma is straightforward.  Note that now the ``mass'' depends
upon the direction of propagation as well as polarisation and position.

\subsection{Photon Capture Cross Sections}

Even without a method for performing the radiative transfer it is possible
to investigate how the combination of dispersion and general relativity
can produce polarisation.  This occurs when one polarisation eigenmode (either
the extraordinary or ordinary) is preferentially captured by the black hole
due to dispersive plasma effects.
This can be quantified by considering the photon capture cross section of
the central black hole for the two different polarisation eigenmodes in the
case of Bondi accretion ($\omega_P^{} \propto r^{-3/4}$ and
$\omega_B^{} \propto r^{-5/4}$). In order to make this
a one-dimensional problem it is necessary to choose an approximation in
regard to the orientation of the wave fourvector relative to the 
external magnetic field.  Here we consider the two extremes,
the quasi-longitudinal ($k^\mu$ parallel to ${\cal B}^\mu$) and the 
quasi-transverse ($k^\mu$ perpendicular to ${\cal B}^\mu$).

\begin{figure}[t!]
\centerline{
\begin{tabular}{cc}
\includegraphics[width=0.5\columnwidth]{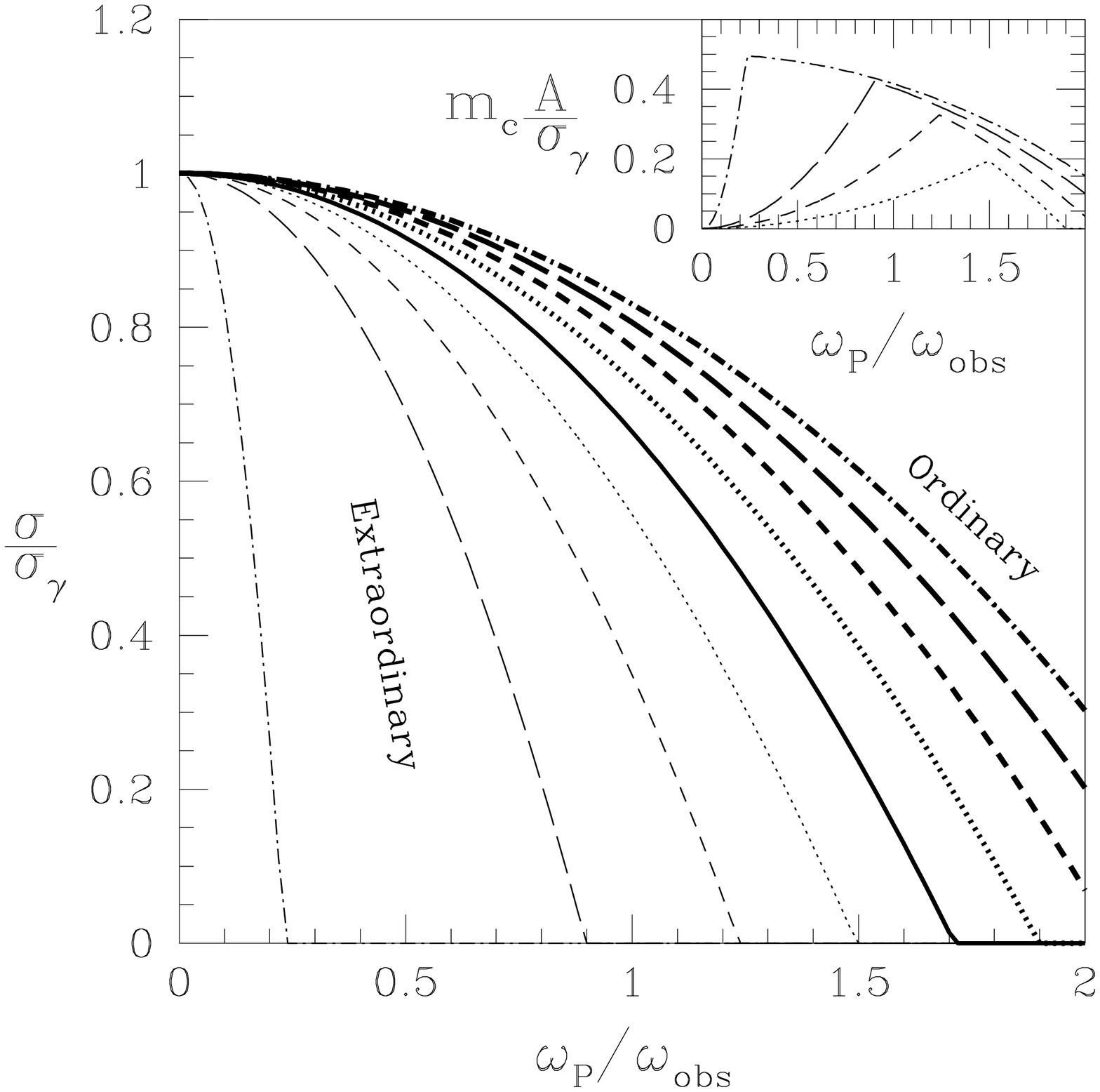} &
\includegraphics[width=0.5\columnwidth]{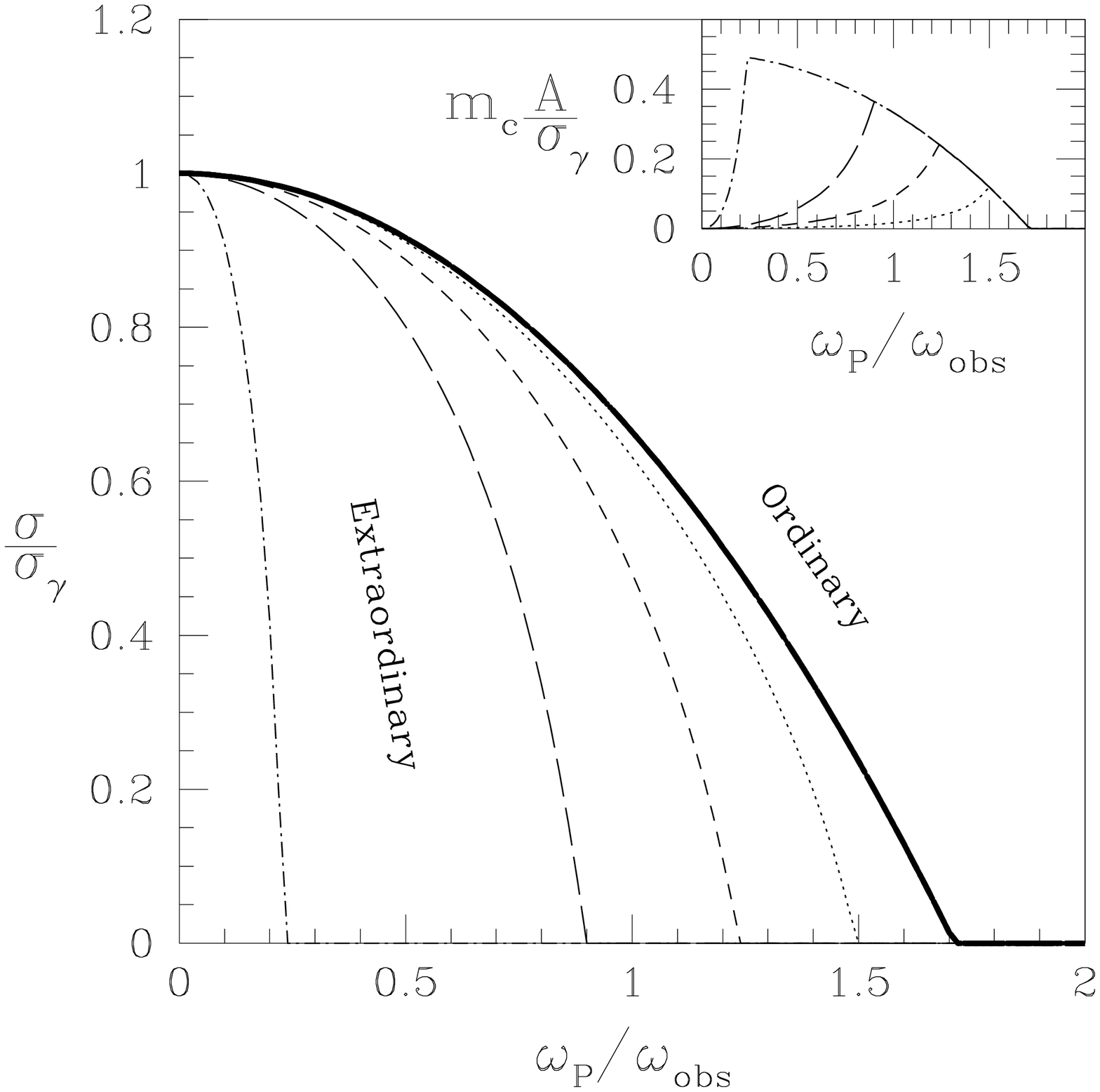} \\
(a) & (b)
\end{tabular}
\caption{Photon capture cross sections in units of the vacuum capture
cross section, $\sigma_{\gamma} = 27 \pi M^2$, for (a) the quasi-longitudinal
and (b) the quasi-transverse approximations as a function of plasma density
for a number of magnetic field strengths.  The solid, dotted, short dashed,
long dashed, and dash-dotted lines correspond to 
$\omega_B/\omega_{\mbox{\tiny obs}} = 0,\,0.7,\,1.4,\,2.1,\,\mbox{and }2.8$, 
respectively.  The insets show the CP fraction, $m_c$ in terms of the
effective emission area $A$ for the same set of magnetic field strengths.}
\label{cross_sections}
}
\end{figure}

As shown in Figure \ref{cross_sections}, in both cases the capture
cross section associated with the extraordinary mode
decreases more rapidly with increasing density than that of
the ordinary mode.  As a result, the black hole will effectively
cast a larger ``shadow'' on the ordinary mode, leading to a net excess
of photons in the extraordinary mode.  If the intervening material is
optically thin, this will lead to an observable net polarisation.  The
magnitude of the polarisation will depend upon the details of the
emission (different regions will have different emissivities) and the
amount of diluting emission from locations far from the black
hole (further than $\sim 5-10 M$).  Both of these
will ultimately depend upon the details of the accretion flow.
However, insight into the second can be obtained by parameterising the
net polarisation in terms of an unknown effective emission area (the
relation of which to the actually emitting area will  still depend
upon the details of the emissivity).  The two extreme cases are shown
in the insets, where the effective emission area is in units
of the vacuum photon capture cross section.  That this will produce much
more CP than LP is a result of the fact that the polarisation
eigenmodes become significantly elliptical only when the angle between
the wave fourvector and the external magnetic field is within $\sim
\omega_P^2 \omega_B^{} / \omega^3$ of $\pi/2$, which is typically
small.

\section{Radiative Transfer}

\subsection{Length Scales \& Radiative Transfer Regimes}

In general the two polarisation eigenmodes will propagate in a coupled fashion.
Because of the dispersive nature of the plasma the general case can be
extremely difficult. Fortunately, it is possible to denote regimes in which
the rays are dispersive and weakly coupled, and non-dispersive and strongly
coupled.

These radiative transfer regimes depend upon two length scales, the plasma
scale length and the coherence length.  The plasma scale length
is the characteristic length scale over which the
plasma changes appreciably,
\begin{equation}
\Lambda_S = \left| \frac{d x^\mu}{d \lambda} \nabla_\mu \ln{n_e} \right|^{-1}
\,,
\end{equation}
written here covariantly, in terms of the affine parameter.  In general this
should also include a measure of the length scales over which the external
magnetic field magnitude and direction change appreciably.  The coherence 
or Faraday length,
\begin{equation}
\Lambda_F = \left| \frac{d x^\mu}{d \lambda}
\left( k_{O \mu} - k_{X \mu} \right) \right|^{-1}
\,,
\end{equation}
is the length over which the two modes will maintain coherence.
The three radiative transfer regimes are then denoted as follows:
\begin{center}
\begin{tabular}{c l}
$\Lambda_F \ll \Lambda_S$ & Adiabatic (weakly coupled \& highly dispersive) \\
$\Lambda_F \simeq \Lambda_S$ & Transitional \\
$\Lambda_F \gg \Lambda_S$ & Nonadiabatic 
	(strongly coupled \& weakly dispersive)
\end{tabular}
\end{center}
Fortunately, the transitional case occurs only for a very small spatial 
region, and can usually be safely ignored.  In these proceedings we have simply
transferred from the adiabatic to the nonadiabatic regimes, skipping
the transitional regime altogether.

The differences between these regimes can be illustrated in context of
Faraday rotation in the interstellar medium.  If in this case the
propagation were adiabatic, and hence the polarisation eigenmodes propagated
independently, the rotation measure would be proportional to
$\int n_e B \, dl$. This is because in the adiabatic regime the extraordinary
mode never evolves into the ordinary mode as a result of changes in the
magnetic field, including field reversals.  However, because in the case of
the interstellar medium the propagation is in reality nonadiabatic, and hence
the modes are strongly coupled, the extraordinary mode can evolve into the
ordinary mode, leading to the familiar rotation measure, proportional to
$\int n_e {\mathbf B} \cdot {\mathbf {dl}}$.

\subsection{Covariance}

Because of the relativistic nature of the problem it is necessary to recast
the radiative transfer in a covariant fashion.  Because the emission and
absorption are local processes, they are most easily dealt with in the
LFCR frame.  It is then necessary to transform from the differential distance
in the LFCR frame, $dl$, into a differential change in the affine
parameter, $d\lambda$. This is accomplished using
\begin{equation}
d l = 
\sqrt{ g_{\mu \nu} \frac{d x^\mu}{d \lambda} \frac{d x^\nu}{d \lambda}
	- \left( u_\mu \frac{d x^\mu}{d \lambda} \right)^2 }
d \lambda \,.
\end{equation} 

In the adiabatic regime the polarisation propagates adiabatically, being
defined by the local plasma conditions, hence only the total intensity need
be transferred.  It is then a simple matter to integrate the occupancy number
instead of the intensity to maintain covariance.

The nonadiabatic regime creates more difficulties as it is now necessary to
propagate a covariant form of the Stokes parameters.  Because each of the
Stokes parameters are defined in terms of intensities and a fiducial
direction, it is possible to define analogous covariant quantities in terms
of the
occupancy numbers and fiducial directions defined by an orthonormal tetrad
propagated along the ray.  Because the ray is no longer strictly a geodesic
this propagation should be done via Fermi-Walker transport
(see \eg \opencite{Misn-Thor-Whee:73}),
\begin{equation}
v^\nu \nabla_\nu e^\mu = \left( v^\mu a^\nu - v^\nu a^\mu \right) e_\nu
\;\; \mbox{where} \;\;
a^\mu \equiv v^\nu \nabla_\nu v^\mu \,.
\end{equation}
However, since
in the nonadiabatic regime the rays are only weakly dispersive, using parallel
transport ($a^\mu = 0$) introduces a negligible error.

\subsection{Emission Models}

We have considered two emission models.  Both are low harmonic
synchrotron emission arising from a power law tail of hot electrons.  The
densities of these electrons were made proportional to the plasma density.
The first model was unpolarised, splitting the emitted power equally between
the polarisation modes.  This was done to better illustrate the creation
of polarisation by the dispersion near the black hole.  The second model
splits the synchrotron flux among the two polarisations appropriately.
This is not necessarily more realistic because in the
accretion flows considered the magnetic field is uniform over the entire
space. While this is not necessary for the dispersive polarisation mechanism
to operate, it will lead to the production of a substantial amount of
polarisation from the synchrotron emission that would not be present
otherwise.

\section{Bondi Flow}

\begin{figure}[t!]
\centerline{
\includegraphics[width=\columnwidth]{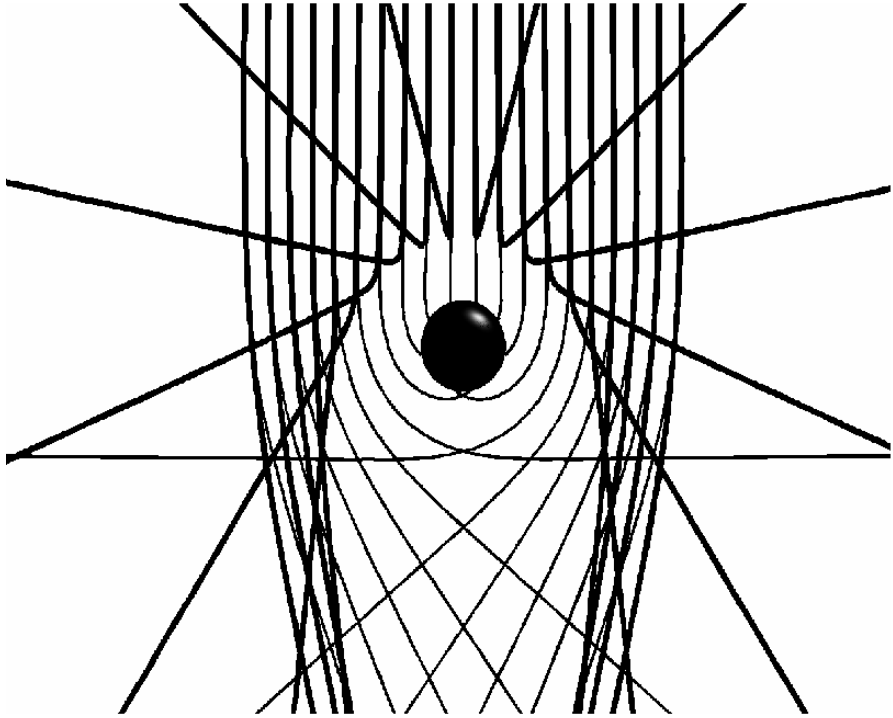}
\caption{ 
A horizontal slice of rays traced through a Bondi accretion flow.  
The ordinary (extraordinary) mode is shown in thin (thick) lines.
}
\label{bondi_rays}
}
\end{figure}

In order to obtain results which can ultimately be compared to observations,
it is necessary to specify the density and velocity of the plasma in the
accretion flow as well as the magnetic field geometry. In general this
should be done in a self consistent manner.  However, for simplicity in 
implementation and clarity of exposition, we have chosen instead to impose a
Bondi accretion flow with a split monopolar magnetic field geometry, the
strength of which is given by a fixed fraction of the equipartition value.
The overall  magnitudes of these parameters are scalable by the observation
frequency. Here, they have been chosen so that interesting effects occur
near 10 GHz, as is the case for spectra of Sgr A$^*$.

It is now possible to explicitly see the dispersion mechanism by tracing
rays, as shown in Figure \ref{bondi_rays}.  Note that the capture cross section
for the ordinary mode is in fact larger than that of the extraordinary mode,
as predicted.

\begin{figure}[t!]
\centerline{
\includegraphics[width=\columnwidth]{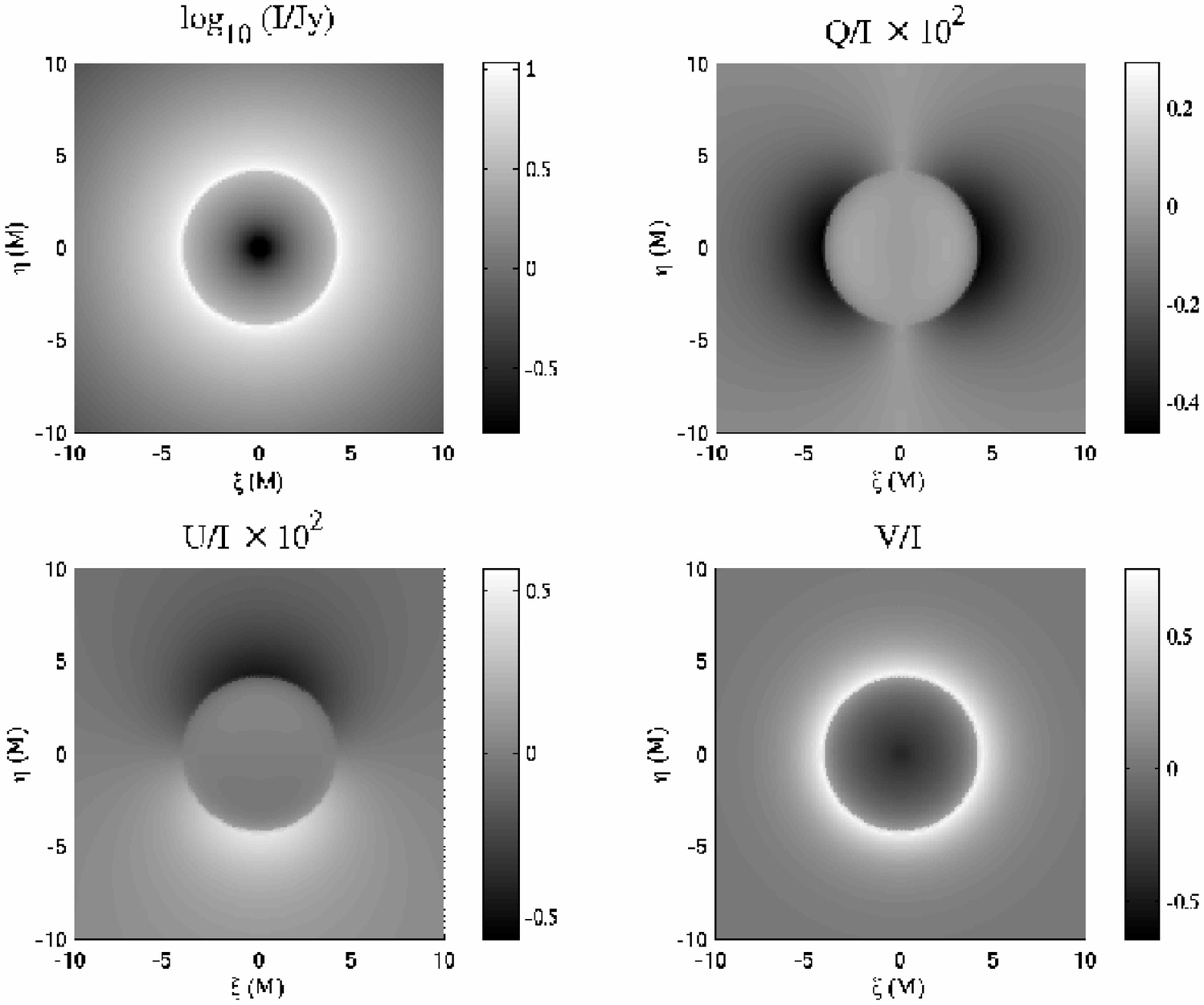}
\caption{Stokes parameters at 10 GHz as observed at infinity as a function of
displacement in the two perpendicular directions ($\xi$ is perpendicular 
to the azimuthal axis) for the unpolarised emission model.  Note that the
scales listed in the titles.
}
\label{iquv1}
}
\end{figure}

\begin{figure}[t!]
\centerline{
\includegraphics[width=\columnwidth]{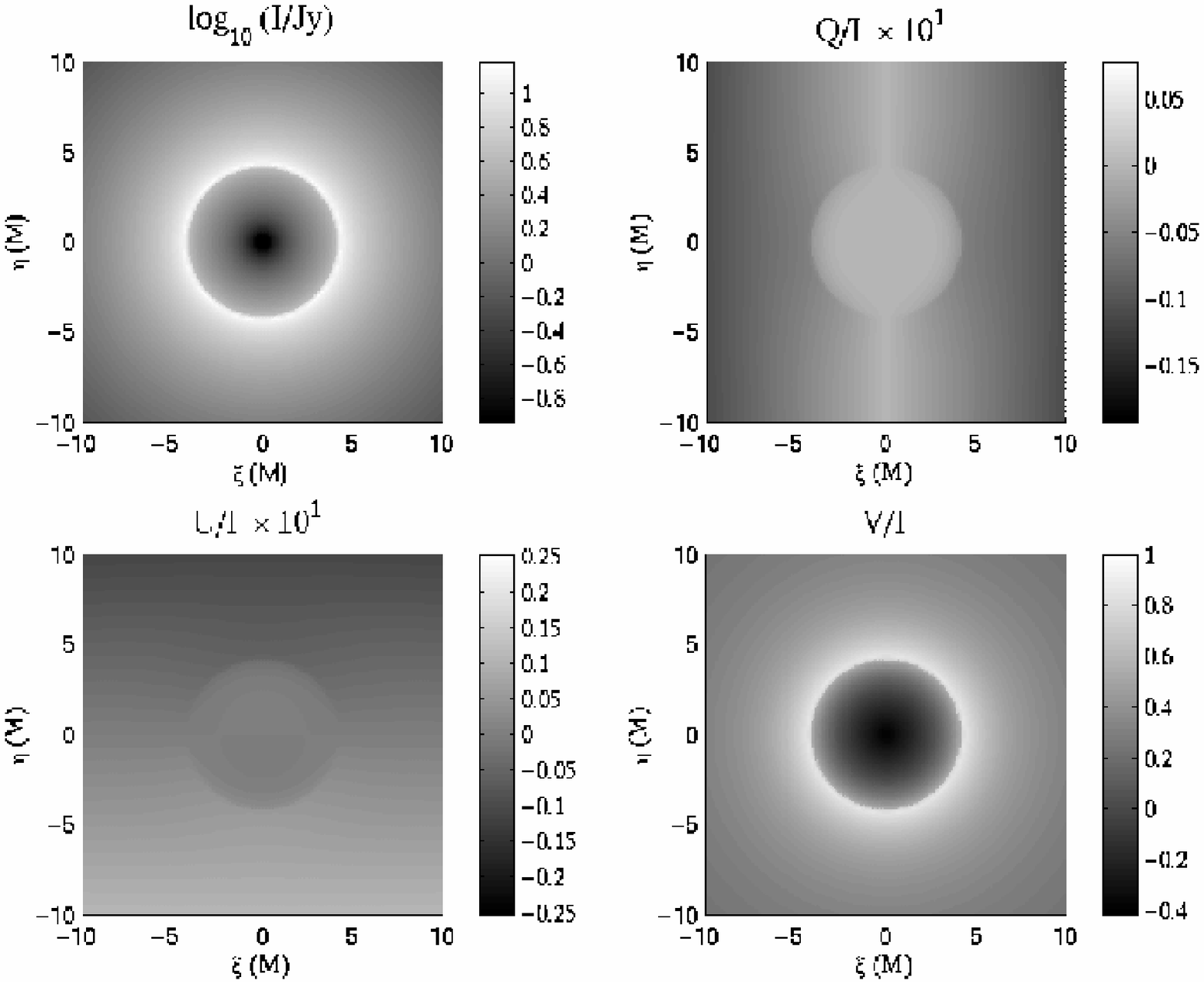}
\caption{Stokes parameters at 10 GHz as observed at infinity as a function
of displacement in the two perpendicular directions ($\xi$ is perpendicular
to the azimuthal axis) for the polarised emission model.  Note that the
scales listed in the titles.
}
\label{iquv2}
}
\end{figure}

The Stokes parameters (Figures \ref{iquv1} and \ref{iquv2}) also
confirm the predictions made
in \S2.3.  While quantitative differences exist, qualitatively the
two emission models produce the similar results.  For both models,
$I \simeq 3$ Jy, $Q/I \simeq -10^{-3}$, and $U/I \simeq -10^{-6}$.  The
disparity between $Q$ and $U$ is a result of the field reversal in the
split monopolar magnetic field geometry occuring in the equatorial plane.
For the unpolarised model, $V/I \simeq 0.2$, and for the
polarised model, $V/I \simeq 0.5$.  All of these numbers are the
integrated values within the regions shown, and therefore do not include
dilution from further out in the accretion flow.  For the unpolarised
model this makes little difference.  For the polarised model, this
means that the LP can be seriously underestimated. However, as mentioned
in \S3.3, this may be an artifact of the artificiality of the accretion
flow geometry at large radii.

\begin{figure}[t!]
\centerline{
\includegraphics[width=\columnwidth]{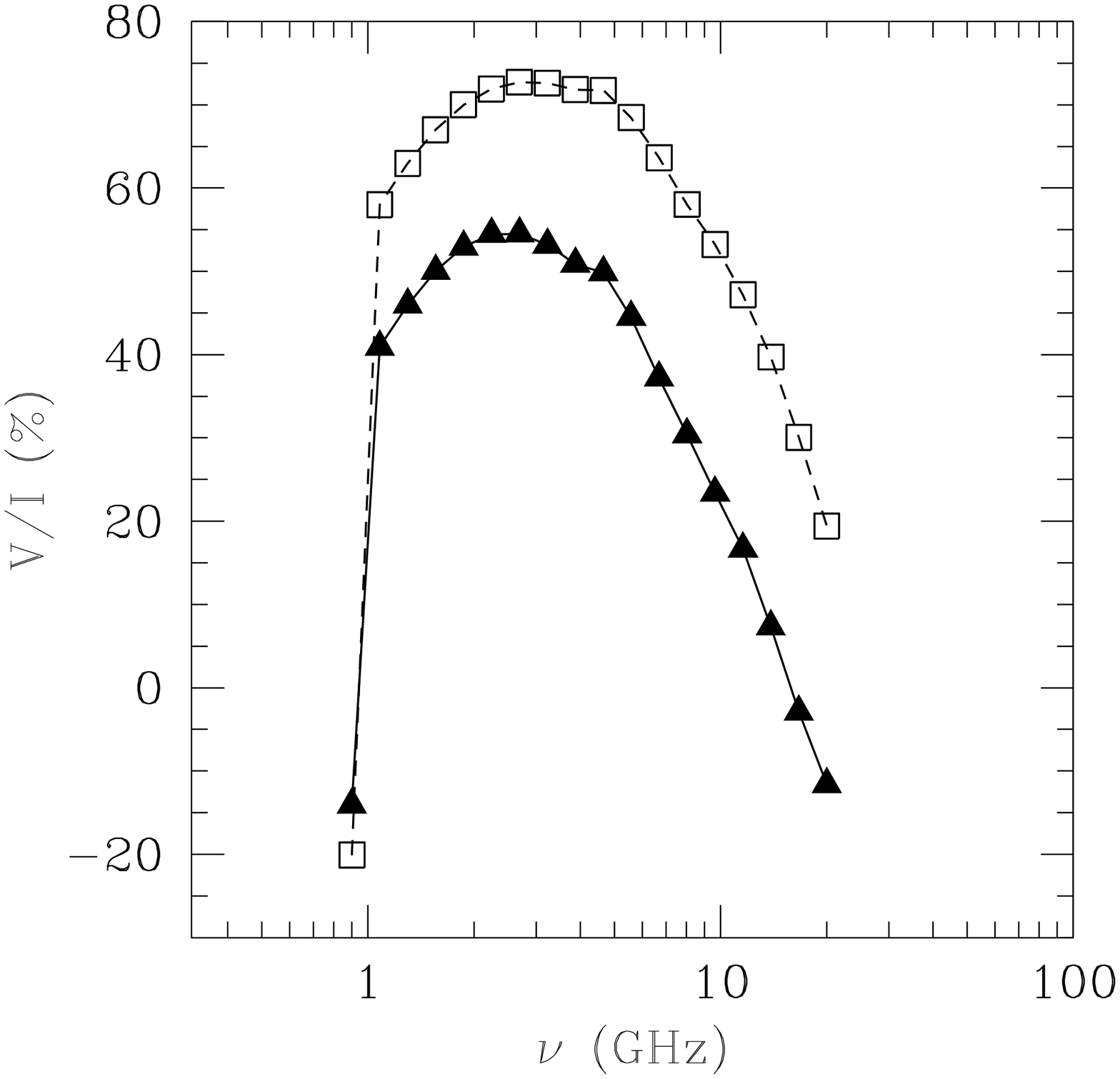}
\caption{The Stokes V parameter as a function of frequency. The filled
triangles (open squares) show V/I for the unpolarised 
(polarised) emission model.}
\label{V_vs_nu}
}
\end{figure}

Plotted as a function of frequency (Figure \ref{V_vs_nu}),
the sizable contribution of the dispersive effects to the total CP
is clearly evident.  At the maximum CP, nearly 75\% of the total
polarisation is due to the dispersive effects alone, as demonstrated by
comparing the curves for the polarised and unpolarised emission models.
Furthermore, the dispersive effects are capable of creating polarisation
over more than a decade in frequency, and hence may be an important source
of polarisation for a significant portion of the spectrum.
This large magnitude is necessary in order to maintain a
significant residual polarisation after the inclusion of diluting
unpolarised emission from the rest of the accretion flow.

\section{Conclusions}

Dispersive effects coupled with general relativistic effects will
produce considerable amounts of CP when the plasma and/or
cyclotron frequencies are commensurate with those being observed.
This method of producing CP is unique in that it does not require a
polarised emission mechanism --- even unpolarised emission will become
polarised after passing near a black hole.  Unlike the
non-dispersive processing mechanisms, \eg Faraday conversion, this
does not require uniform large scale magnetic fields over the entire
disk.  Rather, only uniformity near the black hole horizon is necessary, where
the black hole's influence can in principle moderate the magnetic
field geometry.  This is neither dependent upon the
details of the emission mechanism being employed nor contaminated by large
degrees of LP.

The requirements of the dispersion mechanism
place some constraints upon the emission mechanisms.  The first
is that the mechanism must be able to operate near
$\omega_P^{},\omega_B^{} \sim \omega$.
This can be relaxed somewhat by having the black hole
being backlit, eliminating the necessity for an emission mechanism that is
capable of operating near the hole.
A second constraint upon the emission mechanism
is that it needs to have a large brightness temperature.  This is
equivalent to the fact that the fraction of the total intensity
propagating through the inner $\sim 5-10 M$ must be larger than the CP
fraction. For blazars, this all but rules out
this mechanism (however jets and/or plasma distributions which use dispersion
to magnify the emitting regions may yet make a difference).
For Sgr A$^*$ and M81, brightness temperatures on the order of
$10^{12}$ K are necessary, pushing the upper bounds given by the paucity of
the X-ray fluxes.  Still, this remains a tenable source for the CP in
low luminosity AGN and may be at work in Sgr A$^*$ and/or M81.
Recent LP results (which confirm the earlier observations of 
\opencite{Aitk_etal:00}) suggest that the gas density close
to the black hole in Sgr A$^*$ is far lower than expected for
a conservative accretion flow \cite{Bowe-Wrig-Falc-Back:02}.  Therefore, the Sgr A$^*$ environment is conducive to seeing
relativistic magnetoionic effects at high frequencies close to the black hole.

In addition to applications to low luminosity AGN, this mechanism can
have implications in stellar mass black hole systems as well.  The degree
of CP depends upon the relative sizes of the black hole and the
accretion flow.  Hence, high mass X-ray binaries can be expected
to exhibit a significant amount of CP if ({\em i}) the accretion flow is
optically thin at radio frequencies (\eg if the black hole is being seen
through the corona) and ({\em ii}) magnetic fields are present, presumably
generated via the magnetorotational instability and ordered by the black
hole.

\acknowledgements
The authors would like thank Eric Agol for a number of useful conversations
and comments regarding this work.  This has been supported by NASA grants
5-2837 and 5-12032.

\bibliographystyle{klunamed.bst}
\bibliography{amsterdam.bib}
\end{article}
\end{document}